\documentclass[12pt]{article}
\usepackage{subfig}
\usepackage{natbib}
\usepackage{graphicx}
\usepackage[english]{babel}
\usepackage[utf8x]{inputenc}
\usepackage{amsmath}
\usepackage{graphicx}
\usepackage{tcolorbox}
\usepackage[colorinlistoftodos]{todonotes}
\usepackage{varwidth}
\usepackage{authblk}
\usepackage{algorithm}
\usepackage{algorithmic}
\usepackage{lmodern}
\usepackage{latexsym}
\usepackage{multirow}
\usepackage{amssymb,amsbsy}
\usepackage{mathabx}
\usepackage{amsfonts}
\usepackage{eucal}
\usepackage{longtable}
\def\berr{\begin{eqnarray}}
\def\err{\end{eqnarray}}

\title{The Curious Case of Argon}

\author[1]{Margarita Safonova\footnote{E-mail: margarita.safonova62@gmail.com}}
\author[2]{Alfia Saini}

\affil[1]{Indian Institute of Astrophysics (IIA), Bangalore, India}
\affil[2]{Centre for Interdisciplinary Sciences, Tata Institute of Fundamental Research (TIFR), Hyderabad, India}

\date{}

\begin{document}

\maketitle

\begin{abstract}
In the modern search for life elsewhere in the Universe, we are broadly looking for the following: the planets similar to Earth -- physical indicators of habitability, and the manifestation of life -- the biological signatures. A  biosignature is a measured parameter that has a high probability of being caused by the living organisms, either atmospheric gas species or some surface features. Therefore, the focus of a search is on a product or phenomena produced by the living systems, mostly by microorganisms as these are the most abundant on our planet like, say, methane. However, we may need to distinguish the terms ‘biosignature’ and ‘bioindicator’. A biosignature is what living organisms produce -- a bioproduct, while a bioindicator may be anything necessary for life as we know it, such as water or a rocky planet. Oxygen in this case is a double biomarker; first, it is a byproduct of oxygenic photosynthesis and, second, it is a signature of a complex life, because complex highly organized life requires high levels of oxygen. It is possible that there are other such bioindicators. For example, in the atmospheric compositions of terrestrial planets in our Solar System (including Titan), argon is one of the major constituents, moreover it was recently acknowledged to be a `biologically' active gas, exhibiting organprotective and neuroprotective properties, especially under hypoxic conditions. Here we propose that argon in the atmosphere of a rocky planet is a bioindicator of a highly organized life, provided that the planet is already deemed potentially habitable: with water, atmosphere, and of a certain age allowing for the complex life to evolve. We also delineate its possible detection methods. 

\end{abstract}

\section{Introduction}

In our quest for exo, "other" life in the Universe, different approaches have been applied. In ancient times, people believed that all of the Universe was inhabited. The {\em Mahapuranas}, ancient texts of Hinduism, talk about travelling to other worlds in ‘bodily form’. Ancient Greeks, such as Thales of Miletus and Pythagoras, argued for the plurality of inhabited worlds and, in the beginning of the 20th century, people believed that all the planets of the Solar System were inhabited. Mars was studied by comparing the results of the reflectance spectra of terrestrial plants to the varying colours on its surface, resulting in the birth of astrobotany (Tikhov, 1947). The modern search for inhabited planets began with the search for Earth-like exoplanets: terrestrial mass/radius planets located in a star's habitable zone -- at the right distance from the star to have liquid water on the surface -- just like the Earth (e.g. Schulze-Makuch et al. 2011). But since we cannot penetrate down to the surface of a planet with our telescopes, the concept of a biosignature -- anything that can irrefutably indicate presence of life -- was developed (Des Marais et al., 2002). The most detectable of these are gases in the atmosphere that are produced by living organisms (see e.g. a review on biosignature gases by Seager et al. 2012). For example, methane on Earth is a biosignature -- 90 to 95\% of methane in our atmosphere is biological in origin. However,  methane can also be produced abiotically by various geological processes, and is a major atmospheric constituent on planets non-habitable for us, such as Jupiter or Titan (see Safonova et al. (2020) and refs therein). A recent report on the possible  detection of phospine gas (PH3) in the clouds of Venus \citep{phosphine} generated a lot of excitement, because on Earth it is produced biologically. However, it is also  present in atmospheres of giant planets, though it is supposed to be destroyed in rocky planets' atmospheres or on their surface.

Many other compounds were proposed as biosignatures, for example, water. But the Universe is abound with water, whether in liquid form on the surface (like on Earth), in liquid form under the surface (like on Mars, Enceladus, Europa, Titan, etc., or even Ceres, the largest asteroid in the Solar System), in frozen form in comets and asteroids, or even just in space: the nucleus of IRAS F11506-3851 galaxy (48 Mpc away) contains an amount of water 30 trillion times that of Earth’s oceans combined. In fact, water is one of the most abundant molecules in the interstellar medium (ISM). So, water is rather a bioindicator than a biosignature -- it is an indicator of {\em potential} habitability since water is essential to our kind of life. The Universe is also abound with oxygen, whose abundance rank is 3rd after hydrogen and helium. Along with oxygen, ozone (which protects  the surface from UV radiation) and nitrous oxide (uniquely produced by life on Earth) were suggested as good biosignatures, preferably detected simultaneously. The focus of all commonly considered biosignatures is on a product or phenomena produced by living systems. For example, oxygen is a waste product of oxygenic photosynthetic microorganisms, and most of methane in the atmosphere is a waste product of human activity or biological decay processes. 
 
Though oxygen belongs to the famous CHNOPS\footnote{Elements absolutely necessary for life: carbon, hydrogen (H), nitrogen (N), oxygen (O), phosphorus (P) and sulfur (S)} acronymn, it is a biosignature not only of microbial life, though two-thirds of it in Earth's atmosphere is produced by single-cell organisms. We now know that intelligent life on Earth is only possible due to the large amount of oxygen in our atmosphere. So, oxygen as a gas in the atmospheres of planets is a highly interesting marker -- a double bioindicator; first, it is a byproduct of oxygenic photosynthesis by living systems and, second, a high percentage of oxygen in the air supports an efficient form of living systems' energy metabolism. So, oxygen may also be a signature of complex life, especially on planets that are old enough to allow for that complex life to arise, like the Earth (see, e.g. Safonova et al. 2016). But how complex is complex in this context? Usually, in astrobiological discussions, complex life means macroscopic multicellular life (e.g. Bains \& Schulze-Makuch 2016), but not necessarily highly organized life -- `intelligent' life with brain activity, capable of sophisticated behaviour such as learning, decision making, planning and social interaction, be it bees or humans. It seems that such complex nervous systems require high levels of oxygen \citep{catling}. But it is possible that other gases are also involved.
 
\section{Argon in Planetary Atmospheres}

\begin{figure}
  \begin{minipage}[b]{0.50\linewidth}
\includegraphics[width=1.1\textwidth]{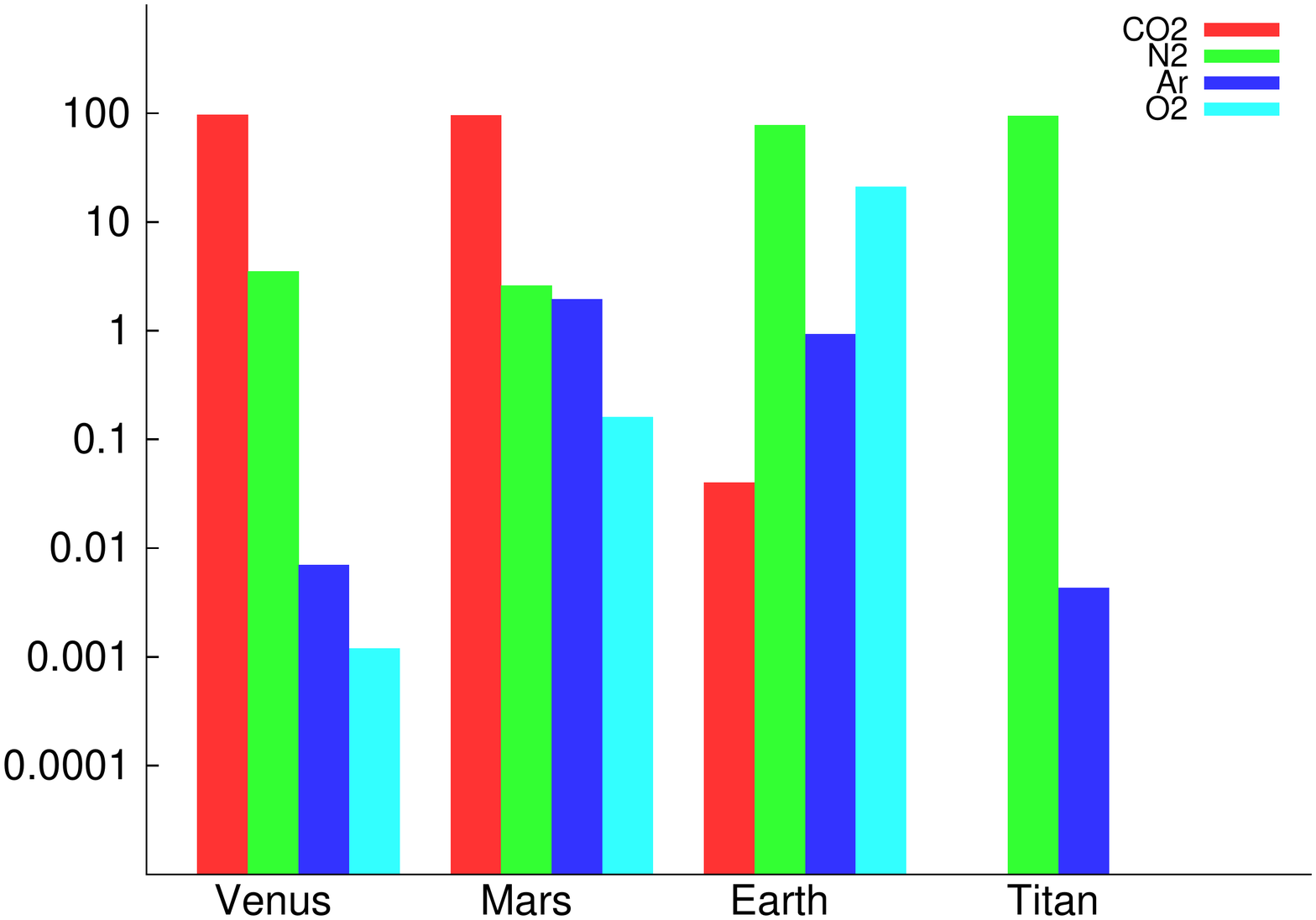} 
    \par\vspace{0pt}
  \end{minipage}%
  \begin{minipage}[b]{0.15\linewidth}
\footnotesize
\begin{tabular}[b]{|ccccc|}
\hline 
\bf Gas (in\%)   &  \bf  Earth   & \bf Venus   & \bf Mars   & \bf  Titan \\    
\hline 
N$_2$   &	78.1  &	3.5  &	2.7  & 94.2  \\
O$_2$   &	21.0  	& 0.003  &	0.15   &  no free O$_2$    \\
Ar   &	0.93  &	0.007  &	1.9   & 0.0043   \\
CO$_2$   &	0.03  &	96 &	95.3  & 0.0000015  \\
\hline
\end{tabular}
    \par\vspace{20pt}
  \end{minipage}
\caption{Major Constituents in the Atmospheres of Terrestrial Planets (by volume).}
\label{fig:argon}
\end{figure}

When we look at the atmospheric composition of terrestrial planets (Fig.~\ref{fig:argon}), we may notice that after the main gases like N2 (Earth, Mars) or CO2 (Venus), argon (Ar) is one of the major constituents: 1.9\% on Mars, 0.9\% on Earth, and 0.007\% on Venus. The amount of argon in Venus's atmosphere is greater than water vapour, and that same phosphine gas: $>110$ ppm of Ar (a sum of three natural isotopes), 20 ppm of water and only 20 ppb of phosphine! On Earth and Mars, Ar is the third dominant component in the atmosphere, and the fourth on Venus. Argon is also one of the major constituents of Titan's surface-level atmosphere after nitrogen, methane and hydrogen, with amounts ranging from 0.005\% (Nieman et al. 2005) to 6\% by volume (Kawai et al. 2013), according to various sources. Titan also possesses ingredients which can form carboxylic acids, piramidines and pyrines (bases of nucleic acids) by the reaction of oxygen from water with N, C, H in tholins. Cassini measurements have shown the existence of a  liquid water ocean under the ice crust -- a potential habitat for life. Mars and Venus were definitely potentially habitable early on, with surface water on Mars at $\sim$3.8 Ga; and lakes, rivers, oceans and even a magnetic field on Venus up to 0.7 Ga \citep{{way},{lammer},{Baines}}. It is interesting that no trace of argon was found on Mercury. 

Argon in the atmosphere exists mostly as three isotopes: the primordial Ar36 and Ar38, the remnants of the initial protosolar cloud, and Ar40, generated by the radioactive decay of K40 present in the crust. The latter is relatively heavy, compared to hydrogen or helium, and thus retains in the atmosphere. 

\section{Argon and its Significance in Life}

In a recent compilation of the {\em biologically} important Mendeleev's periodic table \citep{wackett}, the chemical properties of elements were linked to their biological function. We all know about CHNOPS -- the most essential elements for life, but other elements are also important, like silicon, tin, lead, selenium, etc. The World Wide Web-based University of Minnesota Biocatalysis$/$Biodegradation Database\footnote{UM-BBD, http://eawag-bbd.ethz.ch/index.html} lists the information on microbial interactions with 77 chemical elements because microbial forms of life are the most abundant on our planet. It is notable that argon is listed under the `inert' or `unknown biological function' category (Gao et al. 2010). 

Only quite recently, the biological effects of argon have been getting attention. Initially, the sea divers noted that argon under hyperbaric conditions was exhibiting anaesthetic potency (Behnke \& Yarbrough 1939), though this was not confirmed at normal (normobaric) conditions. It was reported in the 1990s that argon exhibits protective effects on nerve cells (e.g. Soldatov et al. 1998) and provides respiratory resistance under hypoxic conditions, in other words, breathing argon (sometimes up to 55\% in the mixture with only 10\% of oxygen) seems to improve resistance towards hypoxia in animals and humans. In addition, it was later noticed that humans exposed to hypoxia were still able to perform complex manual and mental tasks in an argon-enhanced atmosphere (compared to nitrogen-enhanced) (Antonov \& Ershova 2009). This could be corroborated by a study that showed an increase in oxygen consumption during physical exercise in presence of 30\% argon (Shulagin et al. 2001). Additionally, under a hypoxic atmosphere containing argon, mice exhibited faster development and less teratogenesis, which was attributed to the change of metabolism due to argon (Hollig et al. 2014). It was also shown to be effective in protection against a 1-hr exposure to 85 dB white noise in humans (sounds above 85 are harmful, leading to a permanent hearing loss). 

Argon also protects kidney, heart and lung cells; it has organoprotective properties. Recent study has shown that argon promotes multiple steps in the wound healing process in a diabetic mouse model. It promoted angiogenesis and helped in recruitment of various factors that helped in early recruitment of macrophages and keratinocyte proliferation, thus accelerating wound closure. This is significant in the treatment of diabetic foot ulcers \citep{ning}. However, the most potent function of argon is its neuroprotective function. 

In fact, in 2019 it was finally admitted that argon is a `biologically' active gas (Nespoli et al. 2019). By now, many studies have conclusively demonstrated argon’s cytoprotective and special neuroprotective properties, where argon attenuated brain injury, reduced brain inflammation and preserved neurons from damage due to ischemia (after stroke, Ulbrich \& Goebel 2015). It was pointed out that argon has oxygen-like properties; thus, increase of resistance towards hypoxia may be explained by the same (Höllig et al. 2014). In addition, argon elevates cell survival protein expression against apoptosis. This may be another explanation for its neuroprotective properties. It actually interacts with intracellular signaling molecules in neurons and glial cells. All the studies have shown that argon is especially active in the cells responsible for the higher functions, like microglia, neurons and astrocytes. 

Though research is still scarce on the topic, what evidence has already been presented propels an otherwise benign element into the view of organismal biology.

\section{Hypothesis: Argon as a Bioindicator of `Intelligent' Life}

Till now, the search for biosignatures was directed towards atmospheric gases or surface manifestations that may have been produced by the living organisms. For example, Seager et al. (2013) suggested the classification of biosignature gases on the basis of byproduct gases from metabolic reactions, whether of primary or secondary metabolism. Here, we would like to distinguish the terms `biosignature' and `bioindicator'. A biosignature is what living organisms produce (whether gases in atmosphere or changes on the surface), while a bioindicator may not necessarily be the bioproduct -- it could be anything necessary for life as we know it, such as water or a rocky planet. That is why NASA adopted the strategy "follow the water" in the search for life. So, liquid water on a planet is an indicator of potential habitability -- that life is possible. As we mentioned earlier, oxygen is also a bioindicator; together with the advanced age of a planet it may indicate that complex life has developed. 

In Bains \& Schulze-Makuch (2016), it was concluded that once the complexity arises, intelligence is the   next logical step. But the only `laboratory' we have to derive the results from is Earth. In this article, we are trying to understand what environmental or physical conditions on a planet can facilitate development of intelligence, and how we can infer it. Here, based on the special neuro-related properties of argon described above, we propose that argon in the atmosphere of a rocky planet is a bioindicator of an intelligent life, provided that the planet is already deemed potentially habitable: with water, atmosphere, and of a certain age allowing for the complex life to evolve. On our own planet, it took $\sim$4 billion years for multicellular life to appear, and $\sim$400 millions years more for the cerebrum (responsible for the higher cognitive functions of the brain) to develop. Thus, the planet's age has to be like Earth, at least 4--5 billion years old, if we want to look for `intelligent' life.

\section{Detection Possibilities}

One good thing about atmospheric biosignature gases is that they can be relatively easily detected by space spectroscopic missions. For example, oxygen can be detected in planetary atmospheres by absorption of  stellar light in the IR: at 0.76, 1.06, 1.27 and 6.4 $\mu$m, where the last wavelength is the strongest absorption feature (recently recognized as the only oxygen signature detectable on a terrestrial planet within 5 pc with the upcoming JWST; Fauchez et al. 2020). However, argon on the Solar System planets was detected {\em in situ}, by direct measurements. For example, on Mars it was detected first by the Viking-1 lander in 1976 using a mass spectrometer (Fenselau et al. 2003), and more precise isotopic measurements were done by the quadrupole mass spectrometer (QMS) of the SAM (Sample Analysis at Mars) instrument on Curiosity. 
 
But the future detection of argon in exoplanet  atmospheres has to be done remotely. Argon has many fine structure emission lines at different stages of ionization. When atoms in the planetary atmospheres are subject to electrical discharges (e.g. lightning), they produce a large amount of species, argon itself has more than 300 emission lines. In planetary ionospheres,  the ionization can also occur due to the interaction of a neutral molecule with UV photons, X-rays, or cosmic rays, which are usually ubiquitous and very energetic (up to 100 GeV). 

The emission lines of Ar are not as intense as some of the other strong lines in the optical spectrum but there is a growing number of HII regions for which there are measurements with good signal-to-noise ratio (S/N). This is the case of fine-structure emission line of ArIII at 7136 \AA. It is possible to measure the lines of ArIV at 4713 and 4740 \AA, though it is noted that the former usually appears blended with a line of HeI at 4711 \AA\, that is difficult to correct, so it is better to use the latter. In Pérez-Montero et al. (2007), these lines were used to calculate total Ar  abundances in a sample of HII galaxies from SDSS and a set of HII regions in our Galaxy and the Magellanic Clouds. It is also possible to look for the 1048 \AA\, and 1067 \AA\, resonance fluorescence emission lines of neutral ArI, or the same lines in absorption, using extreme-ultraviolet (EUV) spectrograph; calculations showed that ArI 1067 line would be brighter than the ArI 1048 (Parker et al. 1998). These features may be too weak in the atmospheres of exoplanets for low-resolution space telescopes but future high-resolution spectroscopic missions will be able to detect them.

\end{document}